\documentclass{article}
\usepackage{spconf,amsmath,graphicx}
\usepackage{epstopdf}
\usepackage{float}
\usepackage{subfig}
\usepackage{colortbl}
\usepackage{booktabs}
\usepackage[table,xcdraw]{xcolor}
\usepackage[colorinlistoftodos]{todonotes}
\usepackage{graphicx,wrapfig,lipsum}
\pdfoutput=1

\title{EEG source imaging assists decoding in a face recognition task}
\name{Rasmus S. Andersen$^\ast$,  Anders U. Eliasen$^\ast$, Nicolai Pedersen$^\ast$,\vspace{-5mm}}
\address{\emph{Michael Riis Andersen, Sofie Therese Hansen, Lars Kai Hansen\thanks{This project is supported by the Innovation Fund Denmark project 'Neurotechnology for 24/7 brain state monitoring' and the Novo Nordisk Foundation project 'BASICS'. Authors marked with $^\ast$ made equal contributions.}}\\[2mm]
Technical University of Denmark \\
Department of Applied Mathematics and Computer Science\\
Richard Petersens Plads B324, DK-2800, Kgs.\ Lyngby}
\begin{document}
%
\maketitle
\begin{abstract}
EEG based brain state decoding has numerous applications. State of the art decoding is based on processing of the multivariate sensor space signal, however evidence is mounting that EEG source reconstruction can assist decoding. EEG source imaging leads to high-dimensional representations and rather strong a priori information must be invoked. Recent work by Edelman et al.\ (2016) has demonstrated that introduction of a spatially focal source space representation can improve decoding of motor imagery. In this work we explore the generality of Edelman et al.\ hypothesis by considering decoding of face recognition. This task concerns the differentiation of brain responses to images of faces and scrambled faces and poses a rather difficult decoding problem at the single trial level. We  implement the pipeline using spatially focused features and show that this approach is challenged  and source imaging does not lead to an improved decoding. We design a distributed  pipeline in which the classifier has access to brain wide features which in turn does lead to a 15$\%$ reduction in the error rate using source space features. Hence, our work presents supporting evidence for the hypothesis that source imaging improves decoding.
\end{abstract}
\section{Introduction}
\label{sec:intro}
Brain state decoding based on electroencephalography (EEG) has numerous applications including neurofeedback \cite{arns2014evaluation}, brain computer interfacing \cite{BrainMachine}, and patient monitoring \cite{claassen2013recommendations}.
State of the art decoding is based on signal detection in the multivariate sensor space signal see e.g., \cite{holz2015long}, however evidence is mounting that EEG source reconstruction can assist decoding  \cite{Edelman}.  'Raw' EEG sensor signals are confounded by non-brain signals such as muscle activation, eye blinks and eye motion \cite{BrainMachine} and furthermore  blurred by so-called  volume conduction effects. It is possible that these confounding effects may be suppressed by EEG source imaging.

\begin{figure*}
  \begin{minipage}{1\textwidth}
    \includegraphics[width=1\linewidth]{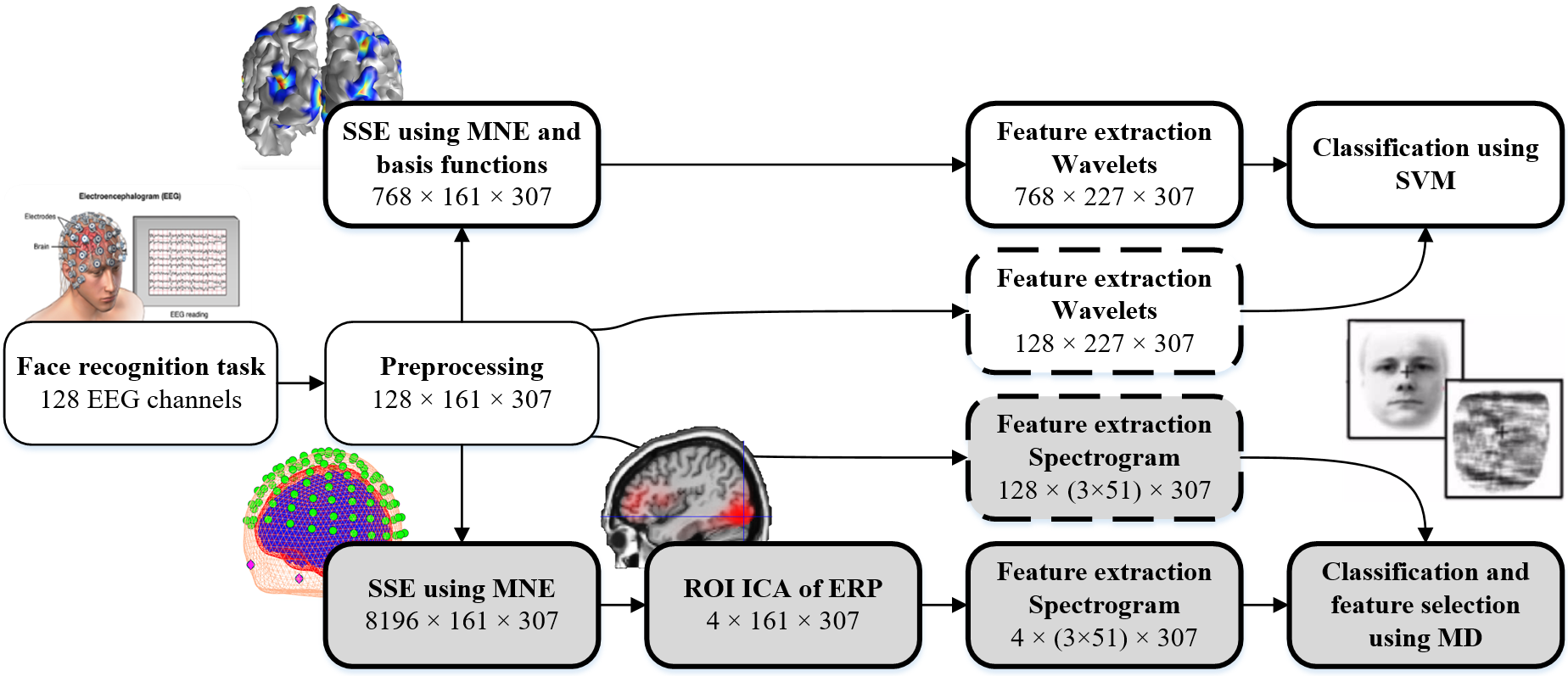}
\end{minipage}
\caption{Flowchart for the investigated pipelines in sensor- and source spaces. Source space estimates (SSE) are obtained using the minimum norm estimator (MNE) \cite{hamalainen1994interpreting}. The grey path illustrates the approach for the focal pipeline, and the white path indicates the process of the distributed pipeline. The dimensions in each square are organized as follows $M, N \times T \times Q$. $M$ and $N$ represent the number of sensor channels and dipoles/basis functions respectively, $T$ is the length of the epoch time series, and $Q$ is the sample size of 'epochs' \cite{SPM8}.}
\label{fig: flow}
\end{figure*}

Inference of  a  high-dimensional EEG source distribution $N=10^3-10^4$, from a relatively low dimensional sensor space measurement $M=10-10^2$, is mathematically ill-posed and requires prior information such as anatomical, functional or mathematical constraints to isolate a unique and plausible solution \cite{baillet2001a}.  The key prior information invoked in Edelman et al.\  is to introduce spatially focused features relevant for detection of motor imagery \cite{Edelman}. The aim of our study is to test the generality of the hypothesis  by applying the approach to a new experimental setting. We have therefore adapted the Edelman et al.\ pipeline to the present task and  denote it the \emph{focal pipeline}, see figure \ref{fig: flow}. It turns out that with this pipeline source imaging does not lead to an improvement in the present data. Instead we suggest to reduce the complexity of the source space representation using a distributed set of spatial basis functions and show that not only does this approach lead to improved decoding error rate it also performs better than a similar pipeline based on raw sensor space signals. The new pipeline is denoted the \emph{distributed pipeline}, see figure \ref{fig: flow}. The rest of the paper is organized as follows: Section 2 presents the focal and distributed pipelines. Section 3 summarizes our results, section 4 provides a discussion of the pipelines and their respective results, and  section 5 presents our conclusions.
\section{Experimental Design and Setup}
\label{sec:setup}
\begin{figure}[H]
\centering
\includegraphics[width=0.5\linewidth]{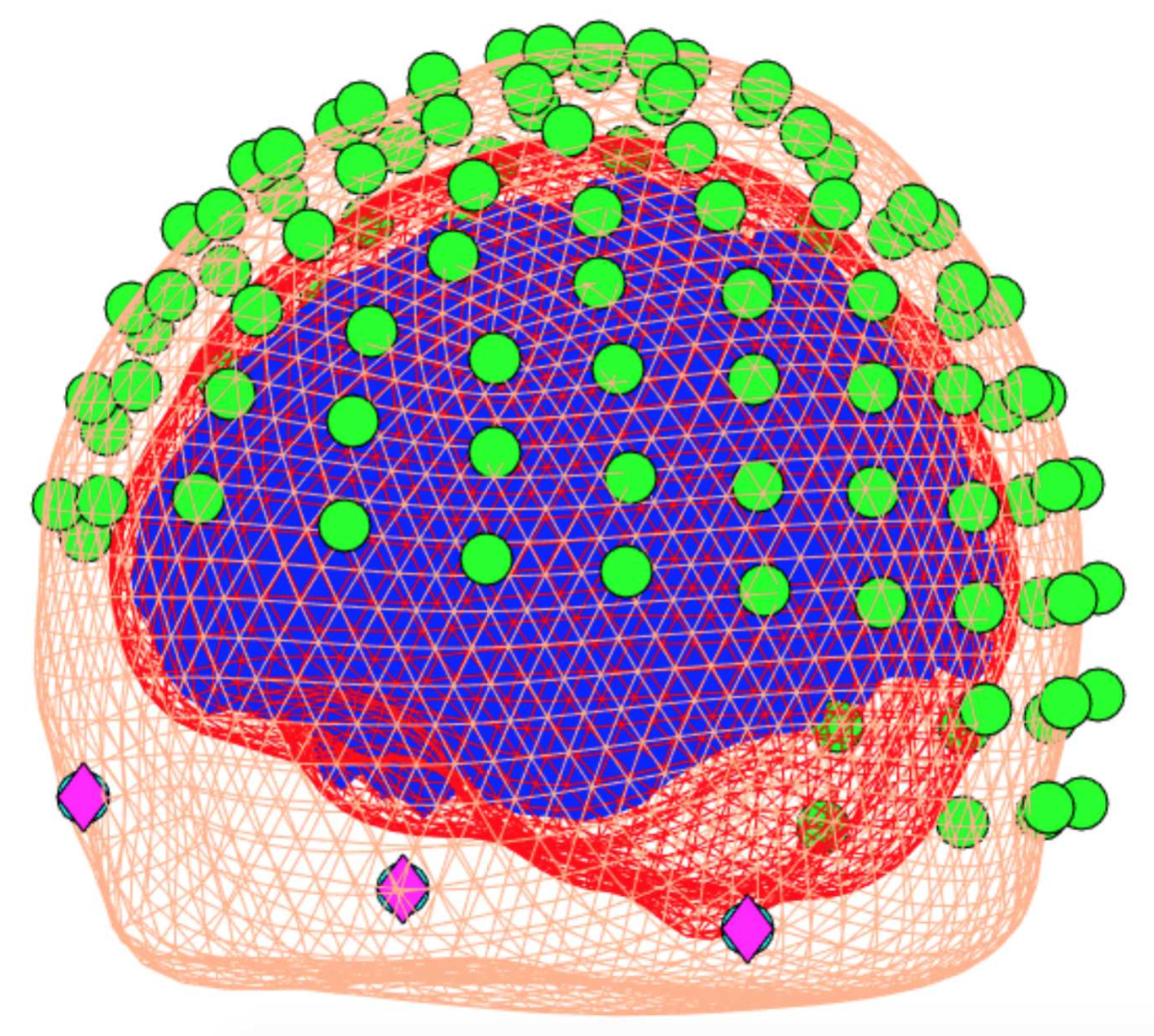}
\caption{Overview of the geometry of source space estimation. The green spheres on the skull represent the EEG electrodes, the purple diamond shaped objects are fiducials (nasion left and right preauricular) for cap alignment. The inner pyramidal mesh illustrates the dipole model of the brain \cite{SPM8}.}
\label{fig: sens2source}
\end{figure}

%
While the study of Edelman et al.\ concerned motor imagery, we here investigate a response known to be distributed with several foci, namely face recognition \cite{henson2011parametric}. The benchmark EEG data set analyzed in this study was collected by D.\ Wakeman and R.\ Henson  and is distributed with the Statistical Parametric Mapping (SPM) toolbox. The data set contains two runs of a subject viewing 86 face images and 86 scrambled face images producing a total of 344 trials.  \cite{henson2011parametric,wakeman2015multi,SPM8}.
%
The Henson and Wakeman  data set was acquired using a 128-channel ActiveTwo system, sampling at 2048 Hz \cite{henson2011parametric,wakeman2015multi}. Two bipolar channels  measuring Horizontal and Vertical Electrooculography (HEOG and VEOG) were removed. The data was preprocessed according to the suggestions from the SPM8 manual \cite{SPM8}. This involved a downsampling to 200 Hz and a re-referencing to 'average reference'. The EEG data was then segmented into epochs of $0.8$ s ($0.2$ s before to $0.6$ s after onset) thus including the entire length of the visual stimuli. Lastly artifact rejection was performed using a threshold approach, which excluded epochs containing channels with amplitudes larger than 200 $\mu V$. The resulting data set contains 307 epochs each consisting of 161 samples, of which 153 are trials with face stimuli and 154 epochs are trials with scrambled face stimuli.
%
Projecting the data from sensor- to source space requires knowledge about the electrical conductance between the scalp sampling sites and the modelled dipole sources of the brain, see figure \ref{fig: sens2source}. Assuming a linear relationship between a given source distribution inside the brain and the electric potentials measured at the scalp \cite{Hauk2004}, this relationship - 'the EEG forward problem' - can be formulated as $\mathbf{B} = \mathbf{A}\mathbf{X} + \mathbf{E}$,
where $\mathbf{B}$ is the $M\times T$ sensor space signal, $\mathbf{A}$ is the $M \times N$ lead field matrix containing the head volume conductor model, $\mathbf{X}$ is the source $N\times T$ space and $\mathbf{E}$ is additive noise.  Because the number of dipoles is typically much larger than the number of electrodes the inverse problem is mathematically ill-posed and strong assumptions are needed to regularize the solution, e.g., of the form $ \min_{X}||\mathbf{A}\mathbf{X}-\mathbf{B}||^2 + \lambda||\mathbf{W}\mathbf{X}||^2,$
where $\lambda$ is a regularization parameter and $\mathbf{W}$ is a spatial matrix that can be designed to promote spatial smoothness of the solution.  The solution is given by $   \hat{\mathbf{X}} = (\mathbf{A}^T\mathbf{A}+\lambda \cdot \mathbf{W})^{-1} \mathbf{A}^{T} \mathbf{B}.$
When $\mathbf{W}$ is the identity matrix, the solution is often referred to as the 'minimum norm estimator' (MNE) \cite{hamalainen1994interpreting}.
%
%
\begin{figure*}[ht]
\begin{minipage}{1\textwidth}
\centering
\subfloat[]{\includegraphics[width=0.27\textwidth]{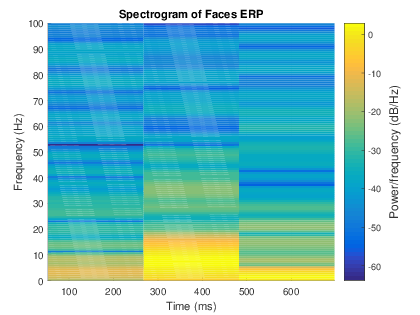}}
\subfloat[]{\includegraphics[width=0.27\textwidth]{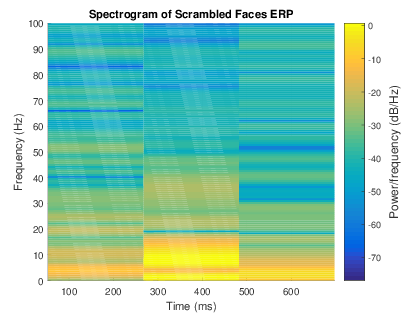}}
\subfloat[]{\includegraphics[width=0.25\textwidth]{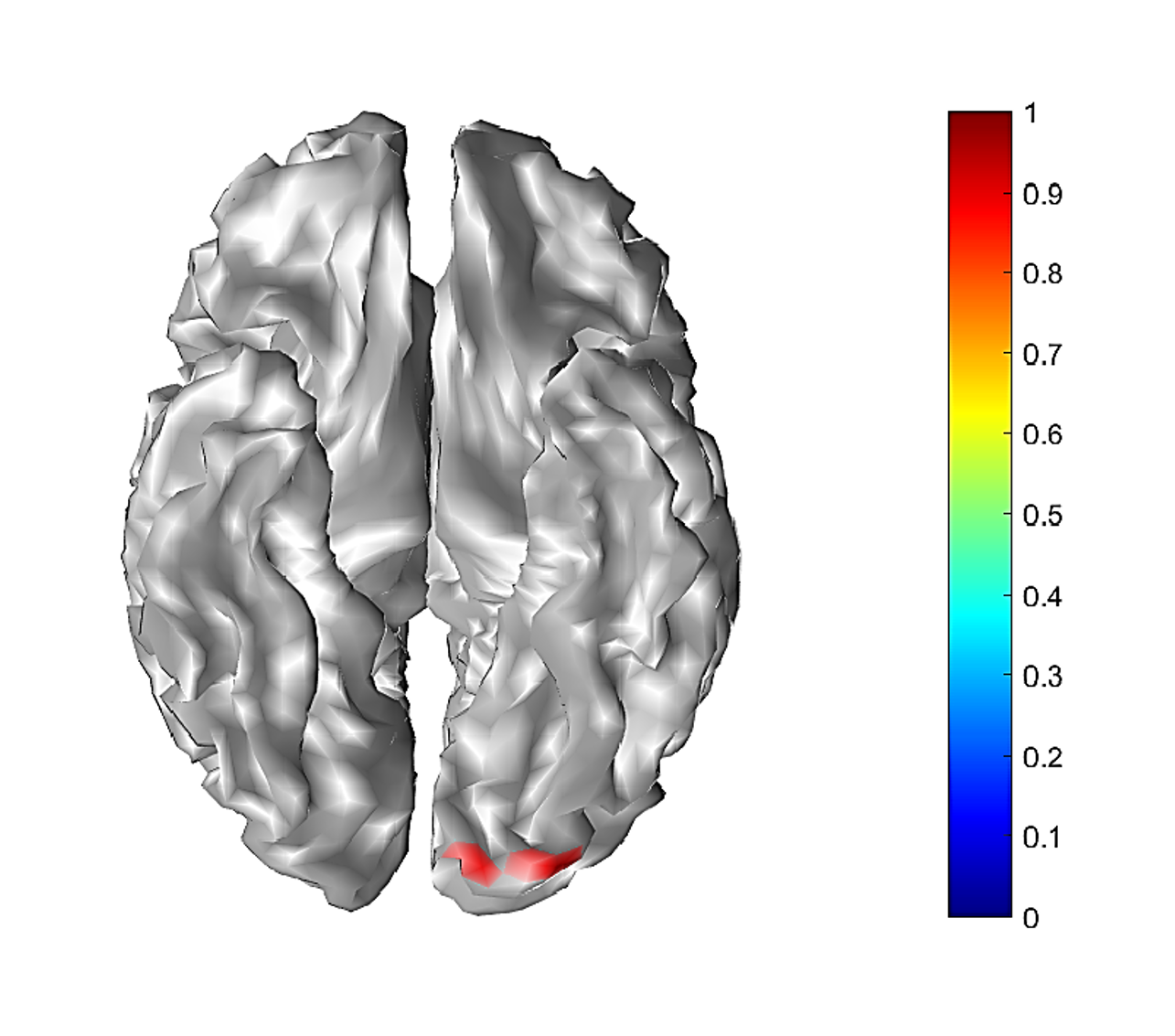}}
\subfloat[]{\includegraphics[width=0.23\textwidth]{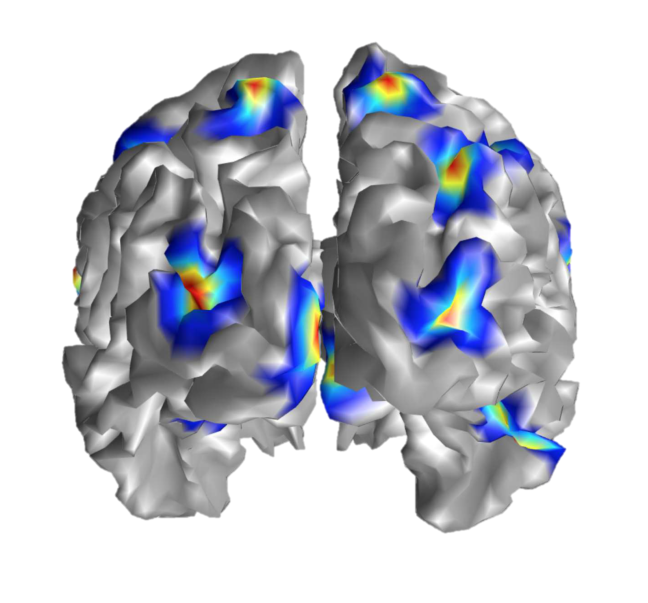}}
\end{minipage}
\caption{(a) shows the spectrogram of the average responses (ERP) for the face trials. (b) spectrogram for the ERP of the trials consisting of scrambled faces. Time and frequency bins of the spectrograms correspond as used in the focal pipeline. (c) is the focal pipeline's ROI scalp map, with the selected dipoles highlighted. (d) displays ten (of 768) basis functions and their distribution across the cortical surface}
\label{Spec}
\end{figure*}
%
%
Following \cite{Edelman} a spatial region of interest (ROI) was derived using a data-driven approach. To identify the ROI for the face recognition task independent component analysis was performed on the event related potential (ERP) for each of the two classes (mean response to faces or scrambled faces) using the extended InfoMax algorithm \cite{InfoMax}.
The independent components were ranked according to how well they separated the two classes based on the difference in their respective spectrogram. The single independent component that displayed the largest separation of the two classes in the time-frequency space was selected. The topography of the selected component was then mapped onto a template brain using the MNE with $\lambda$ tuned by the L-curve method \cite{PCHansen}. The source space representation was thresholded at 75\% of the peak value to identify dipole locations on the cortex to be included in the ROI.
%
%
Also, following \cite{Edelman} the reference sensor space analysis features were extracted from all electrode time series while for the source space analysis time-series from dipoles within the ROI were included. A low resolution time frequency representation of each time-series was obtained based on three time windows and 51 frequency bins in the range of $0-100$ Hz, see figure \ref{Spec}. The most discriminating features were found using the Mahalanobis distance (MD).
Using 10-fold cross validation the data was split into a training- and a test set. A forward feature selection scheme was then applied on the training set to optimize the MD.
\begin{figure}[htb]
\centering
\subfloat[]{\includegraphics[width=0.253\textwidth]{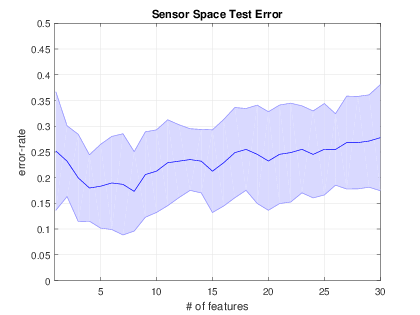}}
\subfloat[]{\includegraphics[width=0.253\textwidth]{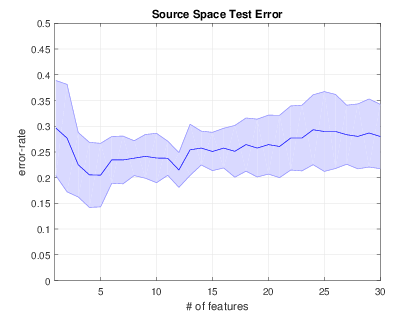}}
\caption{(a) illustrates the result of the feature selection for the sensor space, with the optimal number of features being 8. (b) shows the result of the source space feature selection, where the optimal number of features were 5.}
\label{fig: EdelmanFS}
\end{figure}
Figure \ref{fig: EdelmanFS} illustrates the error rate as a function of the number of features. The lowest error rate in the sensor space was achieved with 8 features, whereas the lowest error rate in source space was found using 5 features.
Having found the optimal feature set, the test set was classified using an epoch based MD-classifier. The mean and covariance matrices were calculated for both classes based on the training set and used to compute the MD to the class means for each test epoch \cite{Edelman}.

For more complex distributed tasks the focal hypothesis may be too restrictive. We therefore suggest an alternative route to simplify the  source space representation. We will apply the basis function approach of Friston et al.\ \cite{priors}, see figure \ref{Spec}(d).  The result of this procedure is a reduced and smoothed representation of the dipole activity. The advantage of this approach is that the entire cortical mesh is spanned by a smaller feature set, in case, a source space representation reduced from 8196 dipole sources to 768 basis functions. We used the MNE estimator with the basis function representation and parameter $\lambda$ was optimized in cross-validation.

Dynamical features were obtained using a wavelet decomposition in the distributed pipeline.  A Daubechies 4 (db4) mother wavelet was used, see e.g., Bahram et al.\ \cite{wave}. Feature extraction was performed for each epoch, viz., ten detail bands were extracted for each of the 768 basis functions and concatenated into one feature vector of 227 entries. All individual feature vectors $\mathbf{f}$ were further combined into a feature matrix, consisting of 174336 features per epoch, i.e., the final data matrix for the distributed pipeline was 174336x307.
%
%
The support vector machine (SVM) is efficient for high-dimensional binary classification tasks \cite{Bishop} and computes
$    y(\mathbf{f}) = \sum_{q=1}^{Q}\alpha_q t_q k(\mathbf{f},\mathbf{f}_q) +b$,
 where $k(\mathbf{f},\mathbf{f}_q) = \mathbf{f}^T \cdot \mathbf{f_q}$  is the linear kernel and $\alpha_q$ are learned SVM parameters for $Q$ data points with targets $t_q = \pm 1$ for face/scrambled face trials. The class label of test observations are obtained as ${\rm sign}(y(\mathbf{f}))$. The performance of the SVM classifier is evaluated using 10-fold cross validation.
\section{Results}
\label{sec:results}
Two different sensor space and two different source space approaches - the focal and the distributed pipelines - were applied for decoding of the face recognition tasks. The results obtained are summarized and compared in table \ref{TableResults}.
\newcommand{\ra}[1]{\renewcommand{\arraystretch}{#1}}
\begin{table}[H]
\centering
\ra{1.8}
\captionsetup{format=plain}
\begin{tabular}{p{2cm} p{2.7cm} p{2.7cm}}\toprule
 \textbf{Pipeline} & \textbf{Sensor space} & \textbf{Source space}\\
\midrule
\rowcolor{black!20} \textbf{Focal} & 17.3\% & 20.5\% \\
\textbf{Distributed} & 10.7\% & 9.1\% \\
\bottomrule
\end{tabular}
\caption{Sensor- and source space test error rate for both pipelines. All the results were obtained using 10-fold cross-validation.}
\label{TableResults}
\end{table}
The focal pipeline achieved an error rate of 17.32\% in the sensor space, using 128 electrodes and the 8 most prominent features extracted from the spectrogram and using the MD classifier. Classification in source space achieved an error rate of 20.48\% using 4 dipole sources and the 5 most prominent features. The source space representation did not improve the classification performance for the focal pipeline for the face recognition task. The distributed pipeline did see a reduced  error rate from 10.7$\%$ in  sensor space to 9.1$\%$ in the source space. The sensor space was comprised of 128 electrodes, where ten db4 detail bands were extracted from each electrode. The source space was spanned by 768 basis functions, also with ten db4 detail bands extracted for each basis function.
\section{Discussion}
\label{sec:discussion}
\begin{figure}[H]
\centering
\subfloat[]{\includegraphics[width=0.25\textwidth]{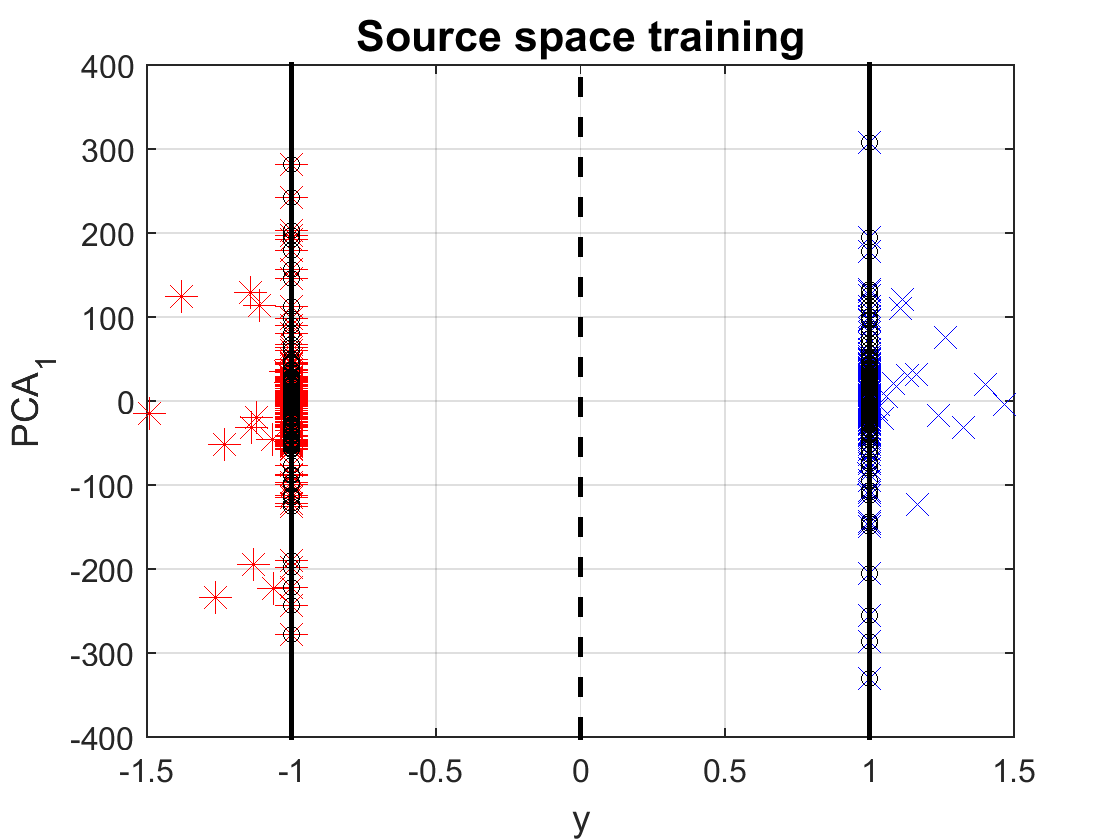}}
\subfloat[]{\includegraphics[width=0.25\textwidth]{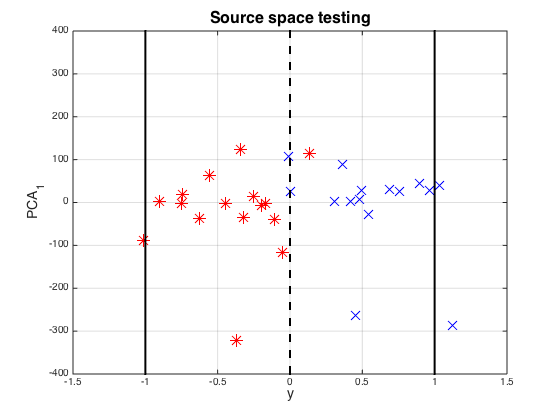}}
\caption{Illustration of the classification problem in the source space. Both plots show one fold in a 10-fold cross-validation. The $x$-axis is the perpendicular distance to the decision boundary and the $y$-axis is the first principal component. (a) shows perfect separation of classes in the training set. (b) shows the classification of the resulting test set. With the present high-dimensional feature representation and relatively limited sample size, \emph{variance inflation} is pronounced in the training set \cite{abrahamsen2012a}, however with the face recognition task's balanced classes this only leads to limited performance loss.}
\label{fig: SVM}
\end{figure}

%
Two different classification pipelines were applied in our study, and for both we compared performances for their sensor space and source space variants. The first approach was based on the \emph{focal} spatial approach proposed in Edelman et al.\ \cite{Edelman}. The pipeline uses a Mahalanobis distance classifier and a small subset of the available dipole sources for classification. Even based on a small number of dipoles feature selection is still critical. Using a forward feature selection the best results were obtained using five features for the source space version and eight features for the sensor space version. This data-driven ROI approach succeeded in detecting the main activity in the occipital region, a region that is known to participate in face recognition \cite{fusiform}.
Our alternative \emph{distributed} pipeline employs a much larger feature space, spanning the entire cortical mesh using 768 basis functions. The distributed pipeline used a linear support vector machine for decoding, which by  margin optimization is somewhat resilient to over-fitting. While the use of the SVM avoided feature selection we did find the typical signature of variance inflation, see figure \ref{fig: SVM}, however due the balanced classes in the present task this has limited effects on the classification performance \cite{abrahamsen2012a}. An analysis of the misclassified epochs in the distributed pipeline showed that  74 \% of the misclassified epochs in the source space representation were also misclassified in the sensor space, indicating that these epochs are indeed hard to decode, presumably due to the generally poor signal to noise conditions of EEG.
%
%
Figure \ref{fig: SVM} shows the training and test classification problem for one of the cross-validation folds in the source space. A perfect separation in the training set is achieved, although the test set is less separated, which is likely due to variance inflation. Despite the variance inflation, which can cause generalization problems for the SVM, the achieved test error of 9.1 \% suggests that the classifier avoids some of these problems in this study.

\section{Conclusion}
Brain state decoding is important to many applications and new evidence was recently put forward by Edelman et al.\ that EEG source imaging can support decoding \cite{Edelman}. The proposed pipeline involves the use of a spatially focal source space representation. The aim of our study was test the generality of this approach. It was not possible to show source space improvements for the decoding in the  face recognition task with the focal pipeline. An alternative distributed pipeline was proposed, which did achieve a  lower  error rate using source decoding reducing the error rate by 15$\%$: from 10.7$\%$ error in the sensor space to 9.1$\%$ in the source space representation. Hence, our result supports the overall hypothesis of Edelman et al.\, namely that a carefully crafted source space representations can improve decoding performance.
\vfill\pagebreak
\bibliographystyle{IEEEbib}
\bibliography{refs}

\end{document}